\documentclass[aps,prl,twocolumn, superscriptaddress, showkeys,longbibliography, 10pt]{revtex4-2}

\usepackage{multirow}
\usepackage{dcolumn}   
\usepackage{bm}        
\usepackage{amssymb}   
\usepackage{graphicx, epsfig, subfigure}
\usepackage[mathlines]{lineno}
\usepackage{color}
\usepackage{enumerate}
\usepackage{enumitem}
\usepackage{tabularx}
\usepackage{amsmath}
\usepackage{tensor}
\usepackage[export]{adjustbox}
\usepackage{upgreek} 
\usepackage{numcompress}

\usepackage{zref-xr} 
\usepackage{zref-user}

\newcommand{\jpsi}{\ensuremath{\mathrm{J}/\psi}}
\newcommand{\psiss}{\ensuremath{\psi(\mathrm{2S})}}

\begin{document}

    \title{Observation of charmonium sequential suppression in heavy-ion collisions at the Relativistic Heavy Ion Collider} 
    \affiliation{Academia Sinica, Nankang, 115, Taipei}
\affiliation{Abilene Christian University, Abilene, Texas   79699}
\affiliation{AGH University of Krakow, FPACS, Cracow 30-059, Poland}
\affiliation{Argonne National Laboratory, Argonne, Illinois 60439}
\affiliation{American University in Cairo, New Cairo 11835, Egypt}
\affiliation{Ball State University, Muncie, Indiana, 47306}
\affiliation{Brookhaven National Laboratory, Upton, New York 11973}
\affiliation{University of Calabria \& INFN-Cosenza, Rende 87036, Italy}
\affiliation{University of California, Berkeley, California 94720}
\affiliation{University of California, Davis, California 95616}
\affiliation{University of California, Los Angeles, California 90095}
\affiliation{University of California, Riverside, California 92521}
\affiliation{Central China Normal University, Wuhan, Hubei 430079 }
\affiliation{University of Illinois at Chicago, Chicago, Illinois 60607}
\affiliation{Chongqing University, Chongqing, 401331}
\affiliation{Creighton University, Omaha, Nebraska 68178}
\affiliation{Czech Technical University in Prague, FNSPE, Prague 115 19, Czech Republic}
\affiliation{Technische Universit\"at Darmstadt, Darmstadt 64289, Germany}
\affiliation{National Institute of Technology Durgapur, Durgapur - 713209, India}
\affiliation{ELTE E\"otv\"os Lor\'and University, Budapest, Hungary H-1117}
\affiliation{Frankfurt Institute for Advanced Studies FIAS, Frankfurt 60438, Germany}
\affiliation{Fudan University, Shanghai, 200433 }
\affiliation{Guangxi Normal University, Guilin, 541004}
\affiliation{University of Heidelberg, Heidelberg 69120, Germany }
\affiliation{University of Houston, Houston, Texas 77204}
\affiliation{Huzhou University, Huzhou, Zhejiang  313000}
\affiliation{Indian Institute of Science Education and Research (IISER), Berhampur 760010 , India}
\affiliation{Indian Institute of Science Education and Research (IISER) Tirupati, Tirupati 517507, India}
\affiliation{Indian Institute Technology, Patna, Bihar 801106, India}
\affiliation{Indiana University, Bloomington, Indiana 47408}
\affiliation{Institute of Modern Physics, Chinese Academy of Sciences, Lanzhou, Gansu 730000 }
\affiliation{University of Jammu, Jammu 180001, India}
\affiliation{Kent State University, Kent, Ohio 44242}
\affiliation{University of Kentucky, Lexington, Kentucky 40506-0055}
\affiliation{Lanzhou University, Lanzhou, 730000}
\affiliation{Lawrence Berkeley National Laboratory, Berkeley, California 94720}
\affiliation{Lehigh University, Bethlehem, Pennsylvania 18015}
\affiliation{Lovely Professional University, Jalandhar - Delhi G.T. Road, Pagwara, Panjab, 144411, India}
\affiliation{Max-Planck-Institut f\"ur Physik, Munich 80805, Germany}
\affiliation{Michigan State University, East Lansing, Michigan 48824}
\affiliation{National Institute of Science Education and Research, HBNI, Jatni 752050, India}
\affiliation{National Cheng Kung University, Tainan 70101 }
\affiliation{Nuclear Physics Institute of the CAS, Rez 250 68, Czech Republic}
\affiliation{The Ohio State University, Columbus, Ohio 43210}
\affiliation{Panjab University, Chandigarh 160014, India}
\affiliation{Purdue University, West Lafayette, Indiana 47907}
\affiliation{Rice University, Houston, Texas 77251}
\affiliation{Rutgers University, Piscataway, New Jersey 08854}
\affiliation{University of Science and Technology of China, Hefei, Anhui 230026}
\affiliation{South China Normal University, Guangzhou, Guangdong 510631}
\affiliation{Sejong University, Seoul, 05006, Korea, Republic Of}
\affiliation{Shandong University, Qingdao, Shandong 266237}
\affiliation{Shanghai Institute of Applied Physics, Chinese Academy of Sciences, Shanghai 201800}
\affiliation{Southern Connecticut State University, New Haven, Connecticut 06515}
\affiliation{State University of New York, Stony Brook, New York 11794}
\affiliation{Instituto de Alta Investigaci\'on, Universidad de Tarapac\'a, Arica 1000000, Chile}
\affiliation{Temple University, Philadelphia, Pennsylvania 19122}
\affiliation{Texas A\&M University, College Station, Texas 77843}
\affiliation{Texas Southern University, Houston, Texas, 77004}
\affiliation{University of Texas, Austin, Texas 78712}
\affiliation{Tsinghua University, Beijing 100084}
\affiliation{University of Tsukuba, Tsukuba, Ibaraki 305-8571, Japan}
\affiliation{University of Chinese Academy of Sciences, Beijing, 101408}
\affiliation{United States Naval Academy, Annapolis, Maryland 21402}
\affiliation{Valparaiso University, Valparaiso, Indiana 46383}
\affiliation{Variable Energy Cyclotron Centre, Kolkata 700064, India}
\affiliation{Warsaw University of Technology, Warsaw 00-661, Poland}
\affiliation{Wayne State University, Detroit, Michigan 48201}
\affiliation{Wuhan University of Science and Technology, Wuhan, Hubei 430065}
\affiliation{Yale University, New Haven, Connecticut 06520}

\author{B.~E.~Aboona}\affiliation{Texas A\&M University, College Station, Texas 77843}
\author{J.~Adam}\affiliation{Czech Technical University in Prague, FNSPE, Prague 115 19, Czech Republic}
\author{L.~Adamczyk}\affiliation{AGH University of Krakow, FPACS, Cracow 30-059, Poland}
\author{I.~Aggarwal}\affiliation{Panjab University, Chandigarh 160014, India}
\author{M.~M.~Aggarwal}\affiliation{Panjab University, Chandigarh 160014, India}
\author{Z.~Ahammed}\affiliation{Variable Energy Cyclotron Centre, Kolkata 700064, India}
\author{A.~K.~Alshammri}\affiliation{Kent State University, Kent, Ohio 44242}
\author{E.~C.~Aschenauer}\affiliation{Brookhaven National Laboratory, Upton, New York 11973}
\author{S.~Aslam}\affiliation{Fudan University, Shanghai, 200433 }
\author{J.~Atchison}\affiliation{Abilene Christian University, Abilene, Texas   79699}
% ...

\collaboration{STAR Collaboration}\noaffiliation

\date{\today}

\begin{abstract}
We report measurements of charmonium sequential suppression in  Ru+Ru and Zr+Zr collisions at $\sqrt{s_{\mathrm {NN}}}$ = 200 GeV with the STAR experiment at the Relativistic Heavy Ion Collider (RHIC). The inclusive yield ratio of $\psi$(2S) to J/$\psi$ as a function of transverse momentum is reported, along with the centrality dependence of the double ratio, defined as the $\psi$(2S) to J/$\psi$ ratio in heavy-ion collisions relative to that in $p$+$p$ collisions. In the 0--80\% centrality class, the double ratio is found to be 0.41 $\pm$ 0.10 (stat) $\pm$ 0.03 (syst) $\pm$ 0.02 (ref), lower than unity with a significance of 5.6 standard deviations. This provides experimental evidence that $\psi$(2S) is significantly more suppressed than J/$\psi$ in heavy-ion collisions at RHIC. This sequential suppression pattern seems to increase from peripheral to central collisions, but with no significant dependence on the transverse momentum.

\end{abstract}


\maketitle



In ultrarelativistic heavy-ion collisions, Quantum Chromodynamics (QCD) predicts the formation of the Quark Gluon Plasma (QGP), in which quarks and gluons are no longer confined within hadrons \cite{Busza:2018rrf,Harris:2023tti}. Charmonia, bound states of charm quarks and their antiquarks ($e.g.$, J/$\psi$ and $\psi$(2S)), are sensitive probes \cite{Prino:2016cni} for studying the QGP, as they are produced predominantly before the QGP formation and experience its evolution subsequently. 

Within the QGP, charmonia are expected to undergo both dissociation and regeneration~\cite{Brambilla:2010cs,Andronic:2024oxz}. The former leads to a suppression of their yields in heavy-ion collisions relative to those in proton-proton ($p$+$p$) collisions~\cite{Matsui:1986dk,Laine:2006ns, Beraudo:2007ky,Brambilla:2008cx,Brambilla:2011sg,Brambilla:2013dpa,Burnier:2015tda,Chen:2017jje}, while the latter partially counterbalances this suppression through the recombination of deconfined charm quarks and antiquarks~\cite{Braun-Munzinger:2000csl,Grandchamp:2003uw,Zhao:2010nk}. Both effects are closely related to the space-time evolution of the QGP and to intrinsic charmonium properties such as mass and size. Recent theoretical advances further relate charmonium dynamics in the QGP to chromoelectric correlators that encode intrinsic transport properties of the QGP~\cite{Binder:2021otw,Scheihing-Hitschfeld:2022xqx,Scheihing-Hitschfeld:2023tuz}.  

Disentangling dissociation and regeneration requires multi-differential measurements, achieved by comparing different quarkonium states, including charmonia and bottomonia (bound states of bottom quark and antiquark), or by varying the QGP space-time evolution through different collision systems and energies. In particular, the $\psi$(2S) meson plays a unique role. Owing to its large spatial extent, approximately 1.8 times that of J/$\psi$~\cite{Satz:2005hx}, $\psi$(2S) probes the QGP at a larger length scale than any other commonly studied quarkonia. For the same reason, $\psi$(2S) production is especially sensitive to the late stages of QGP evolution, as this state is expected to survive only at relatively low temperatures~\cite{Zhao:2010nk,Liu:2009nb,Zhao:2022ggw,Zhao:2022mce}.
In addition, the significantly larger mass of $\psi$(2S) than J/$\psi$ would lead to a suppressed production probability within the framework of statistical models~\cite{Andronic:2019wva}.

Experimentally, $\psi$(2S) production in heavy-ion collisions is commonly studied via its yield or cross section ratio to that of J/$\psi$. This ratio directly probes sequential suppression, wherein the larger and heavier $\psi$(2S) is expected to be more strongly suppressed than J/$\psi$ in the QGP~\cite{Burnier:2015tda,Digal:2001ue,Mocsy:2007jz,Andronic:2019wva}. It also facilitates uncertainty reduction in both experimental measurements and theoretical calculations. The $\psi$(2S) to J/$\psi$ yield ratio has been measured in Pb+Pb collisions at the center-of-mass energy per nucleon-nucleon pair ($\sqrt{s_{_{\rm NN}}}$) of 17.3 GeV \cite{NA50:2006yzz} and 5.02 TeV \cite{CMS:2016wgo,ATLAS:2018hqe,ALICE:2022jeh}. At both energies, the yield ratios in Pb+Pb collisions are smaller than those in $p$+$p$ collisions, confirming the expected stronger suppression for $\psi$(2S) than J/$\psi$. At 17.3 GeV, the ratio suggests a decreasing trend from collisions of small nuclear overlap (peripheral) to those of large overlap (central), consistent with the expectation of increasing QGP effects from peripheral to central events \cite{NA50:2006yzz}. On the other hand, no significant dependence of the yield ratio on nuclear overlap is observed at 5.02 TeV \cite{CMS:2016wgo,ATLAS:2018hqe,ALICE:2022jeh}.

In this letter, the STAR experiment~\cite{STAR:2002eio} at RHIC reports on measurements of inclusive J/$\psi$ and $\psi$(2S) production in Ru+Ru and Zr+Zr collisions at $\sqrt{s_{_{\rm NN}}}$ = 200 GeV. They extend the $\psi$(2S) measurements to a collision system with only half the nucleons of a Pb nucleus and to an energy that bridges the gap between existing results at 17.3 GeV and 5.02 TeV. Both the reduced system size and the intermediate collision energy are expected to influence the space-time evolution of the QGP, thereby modifying the interplay between dissociation and regeneration.

J/$\psi$ and $\psi$(2S) are reconstructed through the electron-positron ($e^+e^-$) channel at mid-rapidity ($\left|y\right| < 1$). In addition to J/$\psi$ and $\psi$(2S) directly produced in partonic scatterings, the inclusive yields also contain decay contributions~\cite{Lansberg:2019adr} from higher charmonium states and hadrons containing bottom quarks, the latter of which is expected to contribute negligibly at 200 GeV though~\cite{Adamczyk:2012ey}. The yield ratio of $\psi$(2S) to J/$\psi$ ($B_{\psiss}\sigma_{\psiss}/B_{\jpsi}\sigma_{\jpsi}$) is presented as a function of charmonium transverse momentum ($p_{\rm T}$), where $\sigma$ represents production cross section and $B$ denotes the branching ratio. Additionally, double ratios, {\it i.e.}, $\psi$(2S) to J/$\psi$ yield ratios in Ru+Ru and Zr+Zr collisions over that in 200 GeV $p$+$p$ collisions, are measured in different intervals of centrality, a quantification of nuclear overlap \cite{Miller:2007ri}.

This analysis utilizes a dataset comprising approximately 1.8 billion $^{96}$Ru+$^{96}$Ru and 1.9 billion $^{96}$Zr+$^{96}$Zr collisions at $\sqrt{s_{_{\rm NN}}}$ = 200 GeV collected in 2018 by the STAR experiment~\cite{STAR:2021mii}. The two collision samples are combined to enhance statistical precision. A minimum bias trigger is used for data recording, which requires coincidence signals from the two Vertex Position Detectors (VPDs) covering the pseudorapidity range of 4.24 $< \left|\eta\right| <$ 5.1 \cite{Llope:2014nva}. The collision centrality classification is determined by matching the charged track multiplicity measured in the Time Projection Chamber (TPC) with the Glauber model \cite{Miller:2007ri}, which also calculates the average number of participants ($\langle N_{\rm part} \rangle$) for centrality intervals~\cite{STAR:2021mii}.

The TPC \cite{TPC2}, Time Of Flight (TOF) \cite{TOF2}, and Barrel Electromagnetic Calorimeter (BEMC) \cite{BEMC}, covering full azimuth and $\left|\eta\right| < $ 1, are utilized for electron identification. Electrons with $p_{\rm T} >$ 0.6 GeV/$c$ are selected based on their normalized ionization energy loss ($n\sigma_{\mathrm{e}}$) measured by the TPC and velocity ($\beta$) derived from TOF. The ratio of energy deposition in the BEMC to the associated track momentum ($E_{0}$/$p$) is applied to further reject hadrons for $p_{\rm T} > 1$ GeV/$c$.

After applying loose initial selections on $n\sigma_{\mathrm{e}}$, $\beta$, and $E_{0}$/$p$, a supervised machine learning technique, based on eXtreme Gradient Boosting (XGBoost) \cite{chen2016xgboost}, is employed to further enhance the charmonia signal significance~\cite{Wang:2025}. 
For the training process, the signal sample is generated via detector simulations using GEANT 3 \cite{Brun:1994aa}, with electrons and positrons from $\psi$(2S) decaying at the collision vertex as inputs. The background sample is selected from $e^+e^-$ pairs in real data whose invariant mass falls within \([2.65, 2.85]\), \([3.25, 3.5]\), and \([3.8, 4.5]\) GeV/$c^2$. After training, the machine learning algorithm assigns a Boosted Decision Tree (BDT) response to each $e^+e^-$ pair, and one can select signal pairs by requiring the BDT response to be above a threshold. The same machine learning model is applied to $\psi$(2S) and J/$\psi$ reconstruction.

\begin{figure}[btph]
\begin{center}
\includegraphics[width=0.9\columnwidth]{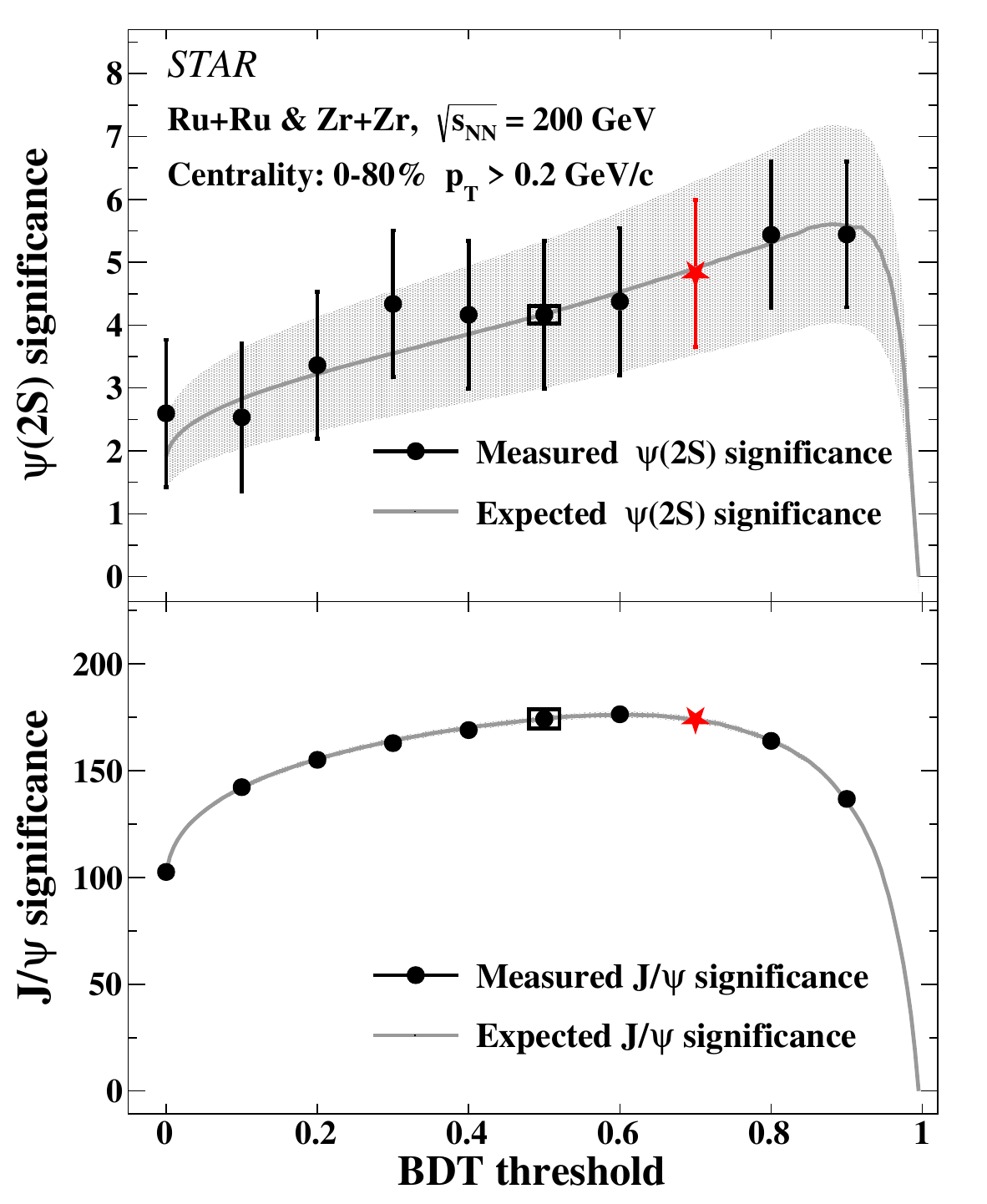}
\caption{Signal significance of $\psi$(2S) and J/$\psi$ as a function of BDT threshold in 0--80\% Ru+Ru and Zr+Zr collisions. The solid markers represent significances extracted from data, while solid curves represent the expected significances from the machine learning. Error bars around data points and bands around solid curves indicate statistical uncertainties. The measured and expected distributions are aligned at the threshold value of 0.5, as indicated by the open boxes. The signal significance corresponding to the default BDT threshold is indicated by stars.}
\label{fig_ML}
\end{center}
\end{figure}

The significances of $\psi$(2S) and J/$\psi$ signals in 0--80\% Ru+Ru and Zr+Zr collisions are shown in Fig.~\ref{fig_ML} as a function of the BDT threshold and compared to those extracted from data. A threshold of 0.5 (marked by the open squares in the figure) is chosen to align the two distributions. The good agreement between the expected and measured significances underscores the accuracy of the machine learning procedure~\cite{Wang:2025}. The dependence of the significance on BDT threshold reflects the interplay of signal efficiency and background rejection rate. Taking into account both signal significance and associated systematic uncertainties, 0.7 is chosen as the default threshold, as indicated by the stars in Fig.~\ref{fig_ML}.

\begin{figure}[btph]
\begin{center}
\includegraphics[width=0.9\columnwidth]{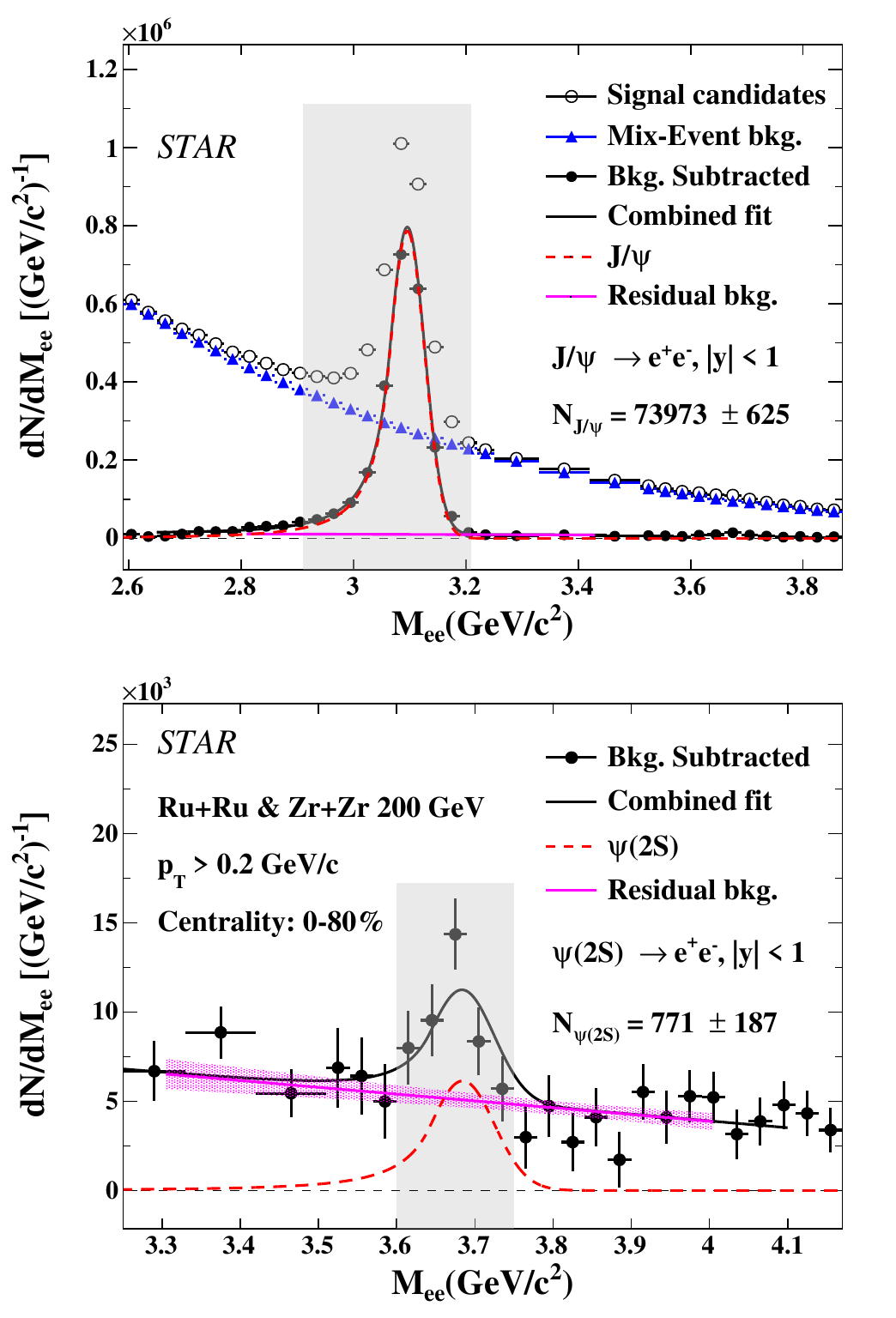}
\caption{Invariant mass distributions of $\psi$(2S) and J/$\psi$ candidates for $p_{\rm T} > 0.2$ GeV/$c$ and $|y| < 1$ in 0--80\% Ru+Ru and Zr+Zr collisions. The combinatorial background estimated using the mixed-event technique is shown as triangles. Background-subtracted distributions (full circles) are fitted. Fit results are shown by the solid curves, while curves of other styles represent individual components included in the fit. The bands around the residual background fit curves display fitting errors. The mass ranges used for determining the signal counts are highlighted as vertical shaded bands.}
\label{fig_SE}
\end{center}
\end{figure}

Figure \ref{fig_SE}, upper panel, presents the invariant mass spectrum of $\psi$(2S) and J/$\psi$ candidates within $|y| < 1$ in 0--80\% Ru+Ru and Zr+Zr collisions at $\sqrt{s_{_{\rm NN}}} = 200$ GeV, using the default BDT threshold. To suppress photon-induced coherent $\psi$(2S) and J/$\psi$ production \cite{PhysRevLett.116.222301,PhysRevLett.123.132302,PhysRevC.97.044910}, a requirement of $p_{\rm T} > 0.2$ GeV/$c$ is applied. The invariant mass distribution above 3.3 GeV/$c^2$ is repeated in the bottom panel. The combinatorial background, estimated using the mixed-event technique \cite{PhysRevC.81.034911}, is depicted as triangles. The filled circles, which show background-subtracted invariant mass distributions, are fitted with two components: a linear function to describe the residual background originating from correlated heavy-flavor decays and Drell-Yan processes, and a Crystal-Ball function, containing a Gaussian core and a power-law tail, to represent the charmonium signal. Parameters of the Crystal-Ball function, except the mean which is fixed according to simulation, are left free in fitting the J/$\psi$ signal. For $\psi$(2S), all the Crystal-Ball function parameters except the magnitude are fixed according to the corresponding parameters for J/$\psi$ extracted from data and ratios of these parameters between J/$\psi$ and $\psi$(2S) from simulation. The raw J/$\psi$ and $\psi$(2S) yields are obtained by counting the number of $e^+e^-$ pairs within the mass ranges of [2.91, 3.21] and [3.6, 3.75] GeV/$c^2$, respectively, and subtracting the residual background contribution based on fitting, taking into account the fitting errors. These raw counts are then corrected for missing signals outside of the mass window, determined using the fitted Crystal-Ball functions.

The $\psi$(2S) and J/$\psi$ reconstruction efficiency ratio is assessed by convoluting the decay kinematics, electron reconstruction efficiency and electron losses due to kinematic requirements ($p_{\rm T} >$ 0.6 GeV/$c$, $\left|\eta\right| < $ 1). For electron reconstruction, the TPC tracking, BEMC matching, and $E_{0}$/$p$ selection efficiencies are assessed with detector simulation, while the TOF matching, $n\sigma_{\mathrm{e}}$ cut and $\beta$ selection efficiencies are evaluated based on a sample of pure electrons from photon conversions in real data.  In addition, the BDT selection efficiency ratio between J/$\psi$ and $\psi$(2S) is determined using BDT response distributions for signal pairs output by the machine learning procedure.

For the reported $\psi$(2S) to J/$\psi$ yield ratios, following systematic uncertainty sources are considered. For the signal extraction process, variations are made regarding the bin width of the invariant mass distribution, the invariant mass ranges used for normalizing the mixed-event distribution and for counting the signal yields, an exponential function used to describe the residual background instead, and the method for calculating raw counts. The root mean square (RMS) of these variations is taken as the signal extraction uncertainty. The uncertainty in the TPC tracking efficiency is estimated by varying TPC-related track quality selections and comparing the differences resulting from these variations. The BDT threshold is scanned from 0.6 to 0.8 in increments of 0.01, and the RMS of these variations is considered the corresponding uncertainty. In the machine learning training process, variations could arise from the randomness in the training, dependency on the initial hyperparameter ranges, and the mass window for selecting the background sample. The RMS of these variations is taken as the systematic uncertainty. Among all sources, the dominant uncertainty stems from the signal extraction process. The total systematic uncertainties for each centrality bin, obtained by summing the individual sources in quadrature, are 12.2\% for 0--20\%, 6.0\% for 20--40\%, 6.2\% for 40--80\%, and 6.3\% for 0--80\% centrality. For the $p_{\rm T}$ differential measurement, the uncertainties are 11.4\% for 0.2--2.0 GeV/$c$ and 4.3\% for 2.0--4.0 GeV/$c$.

To calculate the double ratio, the $p_{\rm T}$-integrated inclusive $\psi$(2S) to J/$\psi$ yield ratio in $p$+$p$ collisions at $\sqrt{s} = 200$ GeV is interpolated from world data \cite{NA50:2006rdp, Clark:1978mg,PHENIX:2011gyb,PHENIX:2019ihw,ALICE:2017leg,ALICE:2014uja,LHCb:2013itw}. Since the world data do not show a significant dependence on charmonium rapidity or collision energy \cite{PHENIX:2019ihw}, a constant function is used to fit the $\psi$(2S) to J/$\psi$ yield ratio as a function of collision energy, using measurements at all rapidity ranges available. The fitted value is adopted, with the fitting error recognized as a source of uncertainty. Additionally, the difference in the interpolated value between the constant function fitting and logarithmic function fitting is considered as another source of uncertainty. The final interpolated $\psi$(2S) to J/$\psi$ yield ratio is $(2.033 \pm 0.085) \times 10^{-2}$. 

\begin{figure}[btph]
\begin{center}
\includegraphics[width=0.93\columnwidth]{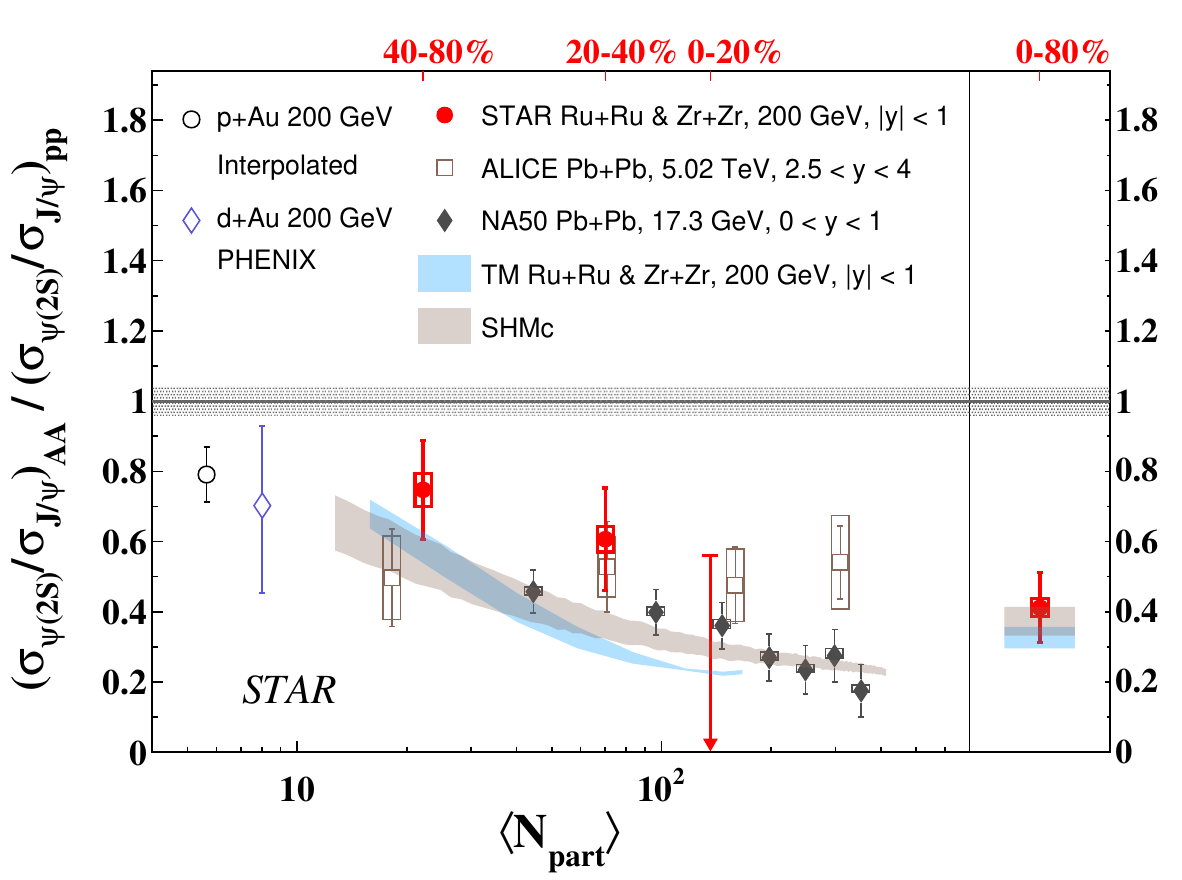}
\caption{Double ratios of $\psi$(2S) to J/$\psi$ as a function of $\langle N_{\rm part}\rangle$ \cite{NA50:2006yzz,ALICE:2022jeh} are shown in the left panel. For the 0--20\% centrality measurements, the upper limit with 95\% confidence level is shown. The result integrated over centrality ($\langle N_{\rm part}\rangle$ = 63) for Ru+Ru and Zr+Zr collisions is presented in the right panel. For measurements from heavy-ion collisions, the error bars indicate statistical uncertainties, while the boxes denote systematic uncertainties. Other results \cite{PHENIX:2013pmn} only display total uncertainties, sum of statistical and systematic uncertainties in quadrature. The bands around unity indicate the uncertainty in the reference $p$+$p$ value.}
\label{fig_phy_cent}
\end{center}
\end{figure}

The Cold Nuclear Matter (CNM) effects, arising from the presence of the nucleus in the collision but unrelated to the QGP formation, can also alter charmonia production in heavy-ion collisions, and be quantified through measurements in $p$+A collisions~\cite{Ferreiro:2009ur,Arleo:2014oha,Gavin:1996yd}. Existing measurements of $p_{\rm T}$-integrated inclusive $\psi$(2S) to J/$\psi$ double ratios are primarily at forward and backward rapidities, and from $p/d/^3\rm{He}$+A collisions at $\sqrt{s_{_{\rm NN}}} = 200$ GeV \cite{PHENIX:2016vmz, PHENIX:2013pmn, PHENIX:2022nrm} and $p$+Pb collisions at $\sqrt{s_{_{\rm NN}}} = 5.02$ TeV \cite{ALICE:2014cgk,LHCb:2016vqr}. Based on the theoretically motivated negative correlation with local particle density \cite{Ferreiro:2009ur,Du:2018wsj}, the $\psi$(2S) to J/$\psi$ double ratio at mid-rapidity in 200 GeV $p$+Au collisions is interpolated to be $0.791 \pm 0.079$, where the local particle density is taken as the average of those at $-2.2 < y < -1.2$ and $1.2 < y < 2.2$~\cite{PHENIX:2016vmz}. It is consistent with the measurement at mid-rapidity in 200 GeV d+Au collisions \cite{PHENIX:2013pmn}, but with significantly reduced uncertainty, as shown in Fig.~\ref{fig_phy_cent}. The interpolated double ratio for $p$+Au collisions can be regarded as an upper limit on the CNM effects in Ru+Ru and Zr+Zr collisions, given possible formation of QGP droplets in $p$+A collisions~\cite{Noronha:2024dtq} and the larger size of the Au nucleus compared to Ru and Zr, which is expected to enhance CNM effects. 

The inclusive $\psi$(2S) to J/$\psi$ double ratio as a function of $\langle N_{\rm part}\rangle$ is shown in the left panel of Fig.~\ref{fig_phy_cent} for Ru+Ru and Zr+Zr collisions at $\sqrt{s_{_{\rm NN}}}$ = 200 GeV. The upper limit with 95\% confidence level for the 0--20\% centrality is determined using the Feldman–Cousins method \cite{Feldman:1997qc}. A hint of decreasing trend from peripheral to central collisions is seen, consistent with the expected increasing hot medium effect towards central collisions. These results are compared to similar measurements in Pb+Pb collisions at $\sqrt{s_{_{\rm NN}}}$ = 17.3 GeV \cite{NA50:2006yzz} and 5.02 TeV \cite{ALICE:2022jeh}. While all three measurements are consistent within uncertainties at comparable $\langle N_{\rm part} \rangle$ values, the double ratio in 17.3 GeV Pb+Pb collisions also shows a decreasing trend from peripheral to central collisions, which is not seen at 5.02 TeV. The observed behavior is likely resulting from the interplay between dissociation and regeneration, both of which are dependent on collision energy.

The double ratio integrated over centrality, shown in the right panel of Fig.~\ref{fig_phy_cent}, is 0.41 $\pm$ 0.10 (stat) $\pm$ 0.03 (syst) $\pm$ 0.02 (ref), where the last term represents the uncertainty in $p$+$p$ reference. It deviates from unity by 5.6 standard deviations ($\sigma$). Since the inclusive J/$\psi$ suppression arises partially from the suppression of $\psi$(2S) that feeds down to J/$\psi$, the double ratio for directly produced $\psi$(2S) and J/$\psi$ is expected to be even smaller than the inclusive double ratio. 
This provides clear evidence that $\psi$(2S) is significantly more suppressed than J/$\psi$ in those heavy-ion collisions at RHIC with respect to $p$+$p$ collisions, consistent with the expected sequential suppression pattern. The double ratio in Ru+Ru and Zr+Zr collisions is also lower than the interpolated value at mid-rapidity for $p$+Au collisions at $\sqrt{s_{_{\rm NN}}}$ = 200 GeV. The significance of the difference is about 3$\sigma$, suggesting the presence of hot medium effects beyond CNM effects. 

The experimental data are also compared to model predictions in Fig.~\ref{fig_phy_cent}. Results for the Tsinghua transport model~\cite{Zhao:2022ggw,Zhao:2022mce,privatecomm}, which incorporates continuous dissociation and regeneration of charmonia in the QGP, are based on parameters presented in Ref.~\cite{Zhao:2022ggw,Zhao:2022mce} and show good consistency with data within uncertainties. The Statistical Hadronization Model (SHMc) \cite{Andronic:2019wva, privatecommSHMc} assumes that all charmonium states completely dissociate in the QGP and regenerate at the hadronization phase boundary. Since it predicts no collision energy dependence, its calculation for 5.02 TeV Pb+Pb collisions at mid-rapidity is compared to this measurement and a reasonable agreement is seen.

\begin{figure}[btph]
\begin{center}
\includegraphics[width=0.93\columnwidth]{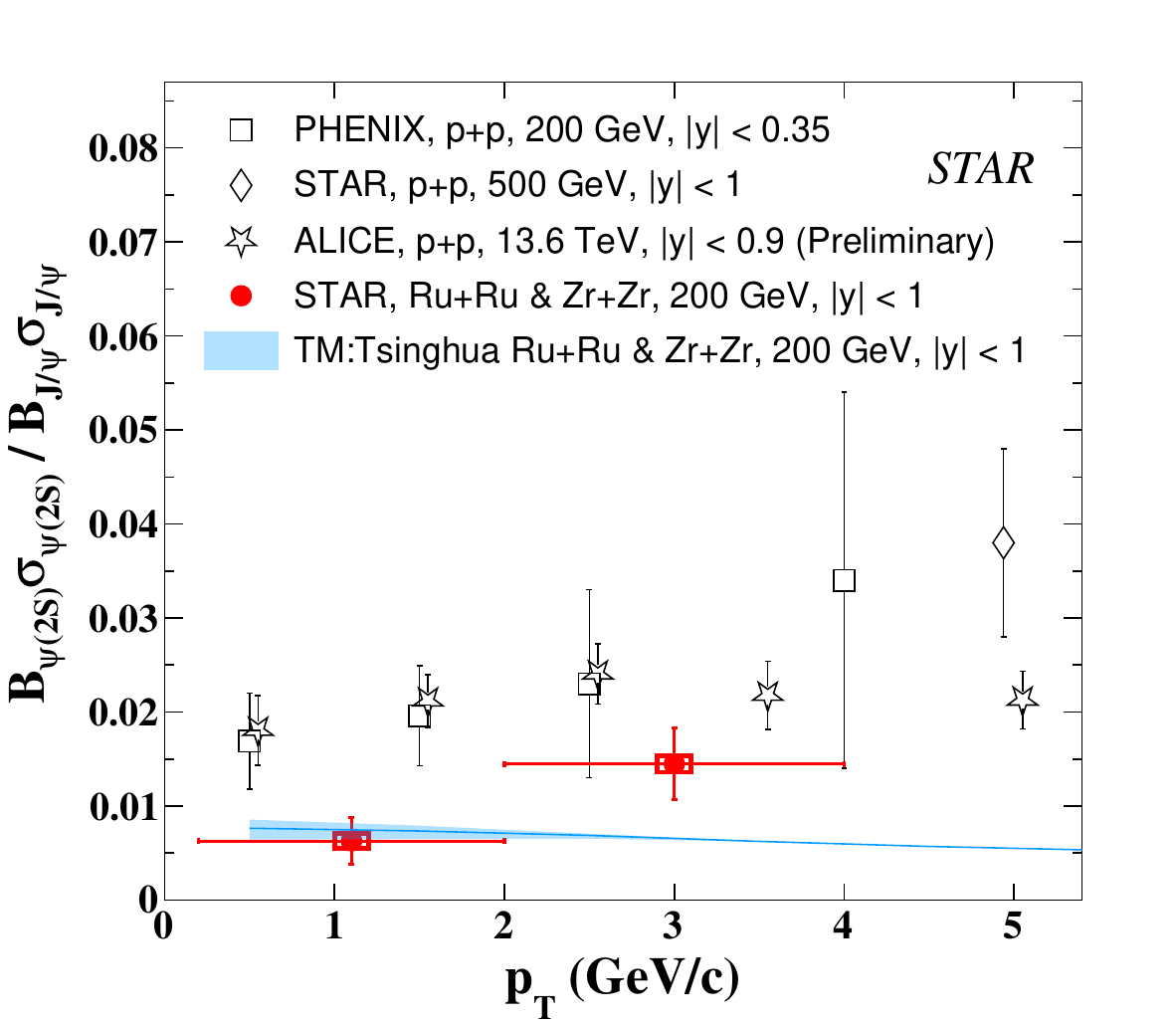}
\caption{Transverse momentum ($p_{\rm T}$) dependence of the $\psi$(2S) to J/$\psi$ yield ratio in $p$+$p$ \cite{PHENIX:2011gyb,STAR:2019vkt, Arsene:2024fdb} and heavy-ion collisions. The vertical bars and boxes around data points represent statistical and systematic uncertainties, respectively, while the horizontal bars indicate bin widths.}
\label{fig_phy_pt}
\end{center}
\end{figure}

Figure~\ref{fig_phy_pt} illustrates inclusive $\psi$(2S) to J/$\psi$ yield ratio as a function of $p_{\rm T}$ in Ru+Ru and Zr+Zr collisions, compared to those in $p$+$p$ collisions \cite{PHENIX:2011gyb,Trzeciak:2014cma, Arsene:2024fdb}. Within $0.2 < p_{\rm T} < 2.0$ GeV/$c$, a constant fit to $p$+$p$ results yields a ratio of $0.0195 \pm 0.0019$, above the Ru+Ru and Zr+Zr result with a significance of 4.1$\sigma$. The significance of the difference is 2.0$\sigma$ for $2.0 < p_{\rm T} < 4.0$ GeV/$c$. The Tsinghua model prediction~\cite{Zhao:2022ggw,Zhao:2022mce,privatecomm} is consistent with data within uncertainties in the 0.2--2.0 GeV/$c$ range, but seems to underestimate data for $2.0 < p_{\rm T} < 4.0$ GeV/$c$, pointing to the need to further improve the description of charmonium interactions with the QGP.

In summary, the STAR experiment at RHIC reports measurements of inclusive $\psi$(2S) and J/$\psi$ production in Ru+Ru and Zr+Zr collisions at $\sqrt{s_{_{\rm NN}}} = 200$ GeV. The $p_{\rm T}$-integrated $\psi$(2S) to J/$\psi$ double ratio in the 0–80\% centrality class is below unity with a significance of 5.6$\sigma$, providing clear evidence that $\psi$(2S) is more suppressed than J/$\psi$ in Ru+Ru and Zr+Zr collisions compared to $p$+$p$ collisions. The double ratio is also smaller than the interpolated value for 200 GeV $p$+Au collisions, suggesting additional suppression of $\psi$(2S) relative to J/$\psi$ in the QGP beyond the CNM effects. A hint of increasing relative suppression from peripheral to central collisions is seen. These results bridge the collision-energy gap in existing $\psi$(2S) measurements in heavy-ion collisions and extend the study to a substantially smaller collision system, thereby providing new constraints on the space-time dependence of charmonium dynamics in the QGP.

We thank the RHIC Operations Group and SDCC at BNL, the NERSC Center at LBNL, and the Open Science Grid consortium for providing resources and support.  This work was supported in part by the Office of Nuclear Physics within the U.S. DOE Office of Science, the U.S. National Science Foundation, National Natural Science Foundation of China, Chinese Academy of Science, the Ministry of Science and Technology of China and the Chinese Ministry of Education, NSTC Taipei, the National Research Foundation of Korea, Czech Science Foundation and Ministry of Education, Youth and Sports of the Czech Republic, Hungarian National Research, Development and Innovation Office, New National Excellency Programme of the Hungarian Ministry of Human Capacities, Department of Atomic Energy and Department of Science and Technology of the Government of India, the National Science Centre and WUT ID-UB of Poland, German Bundesministerium f\"ur Bildung, Wissenschaft, Forschung and Technologie (BMBF), Helmholtz Association, Ministry of Education, Culture, Sports, Science, and Technology (MEXT), Japan Society for the Promotion of Science (JSPS), and Agencia Nacional de Investigacion y Desarrollo de Chile (ANID), Chile.

The data that support the findings of this Letter are openly available~\cite{hepdata.167047}.


\bibliographystyle{apsrev4-2} 
\bibliography{cite}

\end{document}